\documentclass[
	a4paper, 
	10pt, 
	unnumberedsections, 
	twoside, 
]{LTJournalArticle}

\runninghead{} 

\footertext{\textit{}} 

\usepackage{amsmath}

\DeclareMathOperator*{\argminA}{arg\,min} 

\title{A Literature Review on\\Fetus Brain Motion Correction in MRI} 

\author{Haoran Zhang\textsuperscript{1}, Yun Wang\textsuperscript{2}}

\date{\footnotesize\textsuperscript{\textbf{1}}Department of Computer Science, Duke University,\\ \textsuperscript{\textbf{2}}Department of Biomedical Informatics, Emory University}

\renewcommand{\maketitlehookd}{%
	\begin{abstract}
		\noindent This paper provides a comprehensive review of the latest advancements in fetal motion correction in MRI. We delve into various contemporary methodologies and technological advancements aimed at overcoming these challenges. It includes traditional 3D fetal MRI correction methods like Slice to Volume Registration (SVR), deep learning-based techniques such as Convolutional Neural Networks (CNNs), Long Short-Term Memory (LSTM) Networks, Transformers, Generative Adversarial Networks (GANs) and most recent advancements of Diffusion Models. The insights derived from this literature review reflect a thorough understanding of both the technical intricacies and practical implications of fetal motion in MRI studies, offering a reasoned perspective on potential solutions and future improvements in this field. 
	\end{abstract}
}

\begin{document}

\maketitle 


\section{Introduction}

Observing fetal brain development is crucial for diagnosing conditions such as autism and other developmental anomalies \autocite{juul2001prenatal}. Magnetic Resonance Imaging (MRI) offers significant advantages in this context, being precise, rapid, and exhibiting no known harmful effects in either the short or long term when examining brain development \autocite{gowland2011safety}. However, the efficacy of MRI is compromised by challenges such as rapid fetal movements and maternal breathing \autocite{jiang2007mri}. These factors result in relatively low signal-to-noise ratios in the obtained images. Moreover, unlike adults or neonates, fetuses exhibit unrestrained and potentially large movements. Being encased within maternal tissues, the fetal brain occupies a smaller area relative to these surrounding tissues. The above challenges coupled with the low contrast between the fetal brain tissue and surrounding matter, renders imaging interpretation particularly difficult \autocite{rousseau2006registration}.

MRI data acquisition occurs in the Fourier domain, known as k-space, where each data point represents the frequency content of the image. Alterations to even a single point in k-space can impact the entire image. Motion during MRI scans introduces errors into k-space, resulting in blurring and/or ghosting artifacts in the image domain \autocite{hedley1992motion}.

Active research endeavors are underway to mitigate these motion-related issues. Imaging techniques such as Echo Planar Imaging (EPI) \autocite{stehling1991echo} have shown promise in motion reduction by employing specific strategies within EPI. Fast snapshot imaging methods such as Single-Shot Fast Spin Echo (SSFSE) are also commonly used to acquire thick stacks of 2D slices that can largely remove in-plane motion \autocite{saleem2014fetal}.

In recent 15 years, the advancement of Slice to Volume Registration (SVR) algorithms has marked a significant progression towards feasible clinical validation of fetal brain MRI motion correction \autocite{saleem2014fetal, ferrante2017slice, sobotka2022motion, uus2023retrospective, kim2009intersection, ebner2019volumetric, ebner2020automated}. While demonstrated effectiveness, they are still subject to challenges like large motions and initialization failures\autocite{hou20183}. To tackle these challenges, deep-learning based methods have been actively proposed like Convolutional Neural Networks (CNNs) \autocite{hou2017predicting, hou20183, salehi2018real, pei2020anatomy, miao2016cnn}, Long Short-term Memory Network (LSTM) \autocite{singh2020deep}, Transformers \autocite{xu2022svort, xu2023nesvor}, Generative Adversarial Networks (GANs) \autocite{isola2017image, lim2023motion} and Diffusion Models \autocite{xie2022measurement, levac2023accelerated}. To gain a comprehensive understanding of the various methodologies in this field, we have conducted a literature review with a specific focus on algorithmic designs. Our review concentrates on the studies on the fetal brain, intending to not only elucidate the current state of fetal brain MRI motion correction but also offer insights into potential future direction in the field.
\section{Literature Search}

The literature review was conducted using search terms such as "fetal brain," "motion correction," and "MRI." This search encompassed the period from December 15, 2023, to January 20, 2024, focusing on the latest developments in this rapidly evolving field.

In the subsequent sections of this paper, we will methodically examine a variety of methodologies for correcting motion in fetal brain MRI scans. The structure of the presentation is intentionally designed to facilitate a clear and logical understanding, beginning with traditional methodologies and gradually progressing to sophisticated neural network-based approaches. 

These methodologies can be divided into two primary categories: i. Slice-to-Volume Registration, and ii. Deep Learning Methods. Within each category, various studies and techniques are examined. Some of these studies may overlap in terms of their approaches or findings; however, we aim to differentiate them based on their core attributes, such as network features and structures. This progression from foundational techniques to cutting-edge advancements will allow for a comprehensive understanding of the field. Each method will be briefly introduced to provide a contextual framework, paving the way for more in-depth discussion in the following sections.

\section{Slice-to-Volume Registration}

In the past 15 years, numerous studies in the field of SVR have tackled the challenge of fetal motion through retrospective correction \autocite{uus2023retrospective}. \textcite{rousseau2006registration} initially selects a low-resolution MRI stack to establish a global coordinate system, and then aligns each subsequent stack to this reference \autocite{rousseau2006registration}. Following the correction of inter-slice motions, a high-resolution 3D volume is reconstructed using scattered interpolation. \textcite{kim2009intersection} introduced a method for 3D volume reconstruction, which involves co-aligning multiple 2D stacks and employing Gaussian weighting for volumetric reconstruction. \textcite{ebner2020automated} developed a two-step iterative SVR and outlier-robust SRR method, facilitating rapid high-resolution reconstruction in standard space. \textcite{sobotka2022motion} introduced an automated motion correction and volumetric reconstruction framework, subsequently releasing it as an open-source project named NiftyMIC. This framework initially generates a high-resolution reference through Outlier-robust SRR with L2 regularization \autocite{ebner2019volumetric}, then performs slice-wise alignment between the High-Resolution Volume and each stack, culminating in axial stack volumetric reconstruction using Huber L2 regularization.

\begin{figure*} 
	\includegraphics[width=\linewidth]{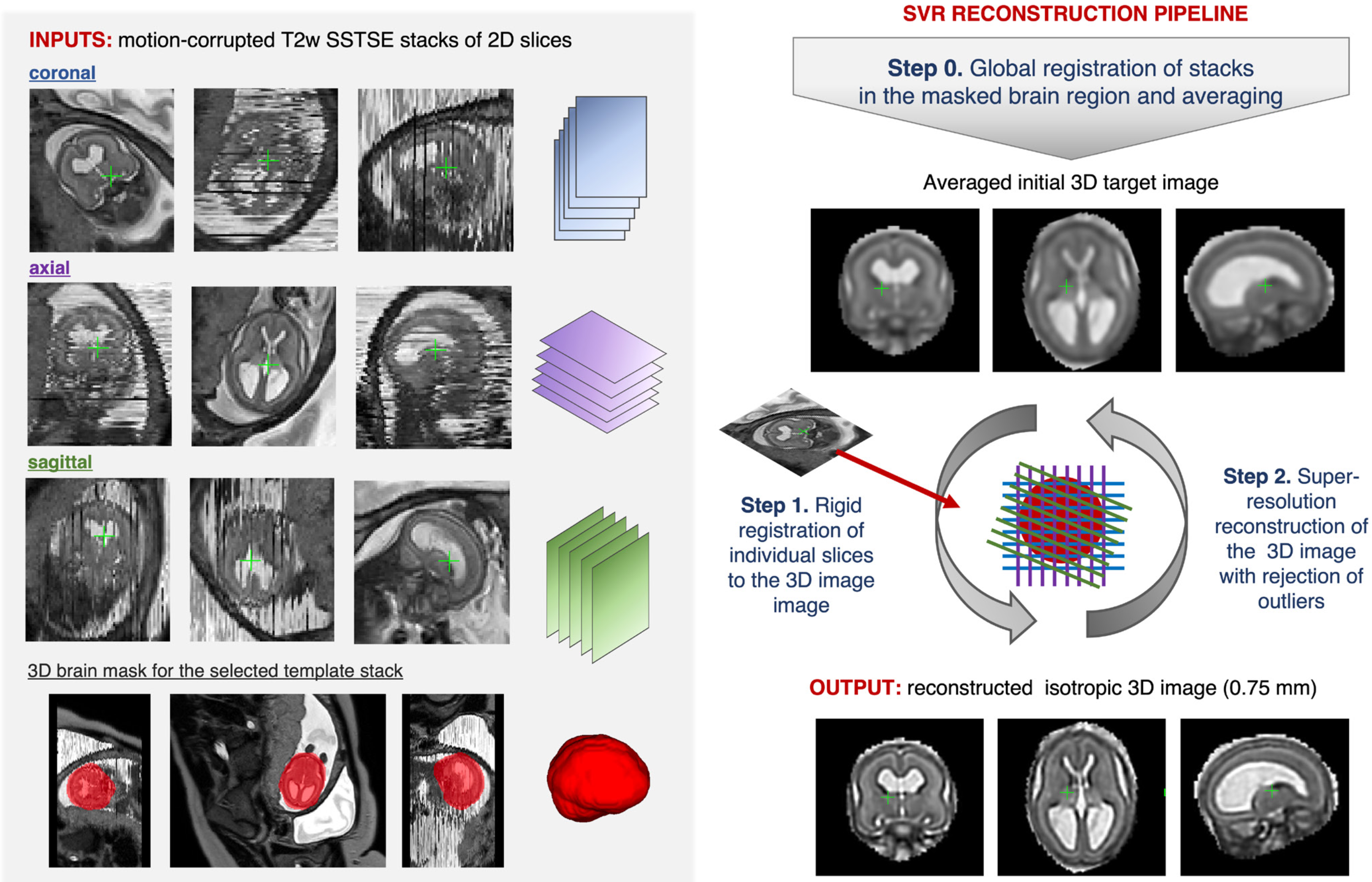}
	\caption{3D SVR reconstruction pipeline for fetal brain MRI. This figure is based on the MRI dataset from St.Thomas’ Hospital, London \autocite{uus2023retrospective}, see the original paper at \href{https://pubmed.ncbi.nlm.nih.gov/35834425}{https://pubmed.ncbi.nlm.nih.gov/35834425}.}
	\label{fig:SVR Flowchart}
\end{figure*}

The pipeline for 3D SVR fetal brain reconstruction is comprehensively depicted in Figure \ref{fig:SVR Flowchart} \autocite{uus2023retrospective}. The process initiates with the global alignment of input image stacks within the designated brain region, establishing a masked area for focused analysis. This initial phase culminates in the creation of a preliminary average 3D image, designated as Step 0 in the methodology.

Subsequently, the procedure progresses to Step 1, where each 2D slice undergoes a process of rigid registration against the current estimate of the 3D image. This step ensures that each slice is accurately aligned in three-dimensional space, contributing to the overall fidelity of the reconstruction.

Following this, Step 2 involves the super-resolution reconstruction of a new, enhanced 3D image. This is achieved by integrating all the aligned slices, thus improving the resolution and clarity of the resulting image.

The methodology adopts an iterative approach, where Steps 1 and 2 are alternated repeatedly. This iterative process is crucial for achieving convergence towards a stable and accurate reconstruction of the fetal brain. Throughout these steps, there is also a quality control mechanism in place. This mechanism involves the exclusion of any slices that are either misaligned or exhibit signs of corruption, ensuring the integrity and reliability of the final reconstructed image \autocite{uus2023retrospective}.

Slice-to-volume registration is defined as the process of aligning a two-dimensional (2D) image $I$ with a three-dimensional (3D) volume $J$. The objective is to determine a transformation function $\hat{\Theta}$ that best aligns the 2D tomographic slice $I$ to the 3D volumetric image $J$. This alignment is achieved by minimizing an objective function as outlined in Equation \ref{eq:svr_obj}:

\begin{equation}
\hat{\Theta} = \argminA_\Theta \mathcal{M}(I, J; \Theta) + \mathcal{R}(\Theta)
\label{eq:svr_obj}
\end{equation}

In the aforementioned equation, $\mathcal{M}$ signifies the image similarity term, commonly known as the matching criterion. This term evaluates the similarity between the 2D image and its corresponding slice in the 3D volume, predominantly based on intensity information or distinctive structural features present in $I$ and $J$. Concurrently, $\mathcal{R}$ symbolizes the regularization term, which applies constraints to guarantee a well-defined solution. This term is particularly pivotal in governing the geometric characteristics of the transformation model, especially crucial in scenarios involving non-rigid registration \autocite{uus2023retrospective}.

The classification of the registration process as either rigid or non-rigid hinges on the permissible deformations in image $I$ or its corresponding reformed slice from $J$. In rigid slice-to-volume registration, the transformation is confined to rigid motions, typically incorporating six degrees of freedom. Conversely, non-rigid registration allows for more intricate transformations, encompassing deformations or advanced linear transformations such as affine transformations \autocite{ferrante2017slice}.

The selection of a regularizer in the function $\mathcal{R}$ is contingent upon the specific transformation model employed. While simpler models, such as rigid body transformations, may be estimated even in the absence of a regularizer, more complex non-rigid models necessitate the inclusion of $\mathcal{R}$ to ensure the realism of the transformations. In the context of slice-to-volume registration, this regularizer can impose planarity constraints (in situations where out-of-plane deformations are restricted) or can limit the extent of out-of-plane deformations to yield plausible outcomes \autocite{ferrante2017slice}. Furthermore, when relevant, incorporating knowledge about the tissue's elasticity into the regularizer can enhance the model's accuracy. The ultimate objective is to optimize the energy as delineated in Eq. (\ref{eq:svr_obj}), identifying the most precise $\hat{\Theta}$ that aligns the 2D and 3D images both effectively and realistically.

While traditional iterative methods for motion correction and reconstruction have shown effectiveness, they are hindered by two major limitations: the inability to accurately estimate large motions, and a heavy reliance on precisely setting initial transformation parameters \autocite{hou20183}. Furthermore, the similarity measures optimized in these intensity-based methods often exhibit a highly non-convex nature. This aspect greatly increases the risk of the optimizer being trapped in local maxima, thus restricting the capture range of these techniques \autocite{miao2016cnn}.

The capture range of SVR is notably limited, primarily because it depends on iteratively optimized intensity-based similarity metrics. These metrics serve merely as proxies for aligning slices to a reference volume, without ensuring accurate alignment. The presence of a motion-free reference volume is not always guaranteed. To broaden the capture range, one can employ strategies like grid search on rotation parameters along with multi-scale registration \autocite{taimouri2015template}. However, these approaches are computationally demanding due to their reliance on iterative numerical optimization in the testing phase. Alternatively, age-matched atlases, such as those referenced in \autocite{gholipour2017normative}, can be used as reference volumes. Atlas-based registration methods, detailed in \autocite{taimouri2015template, tourbier2017automated}, are also possible, but their computational intensity limits their applicability in real-time scenarios.

Accurate motion correction remains a significant challenge in the field. Supervised learning has been explored in 2D/3D registration, with some studies proposing metric learning approaches to develop customized similarity measures through supervised learning \autocite{bronstein2010data, michel2011boosted}.

\section{Deep Learning Methods}
\begin{figure*} 
	\includegraphics[width=\linewidth]{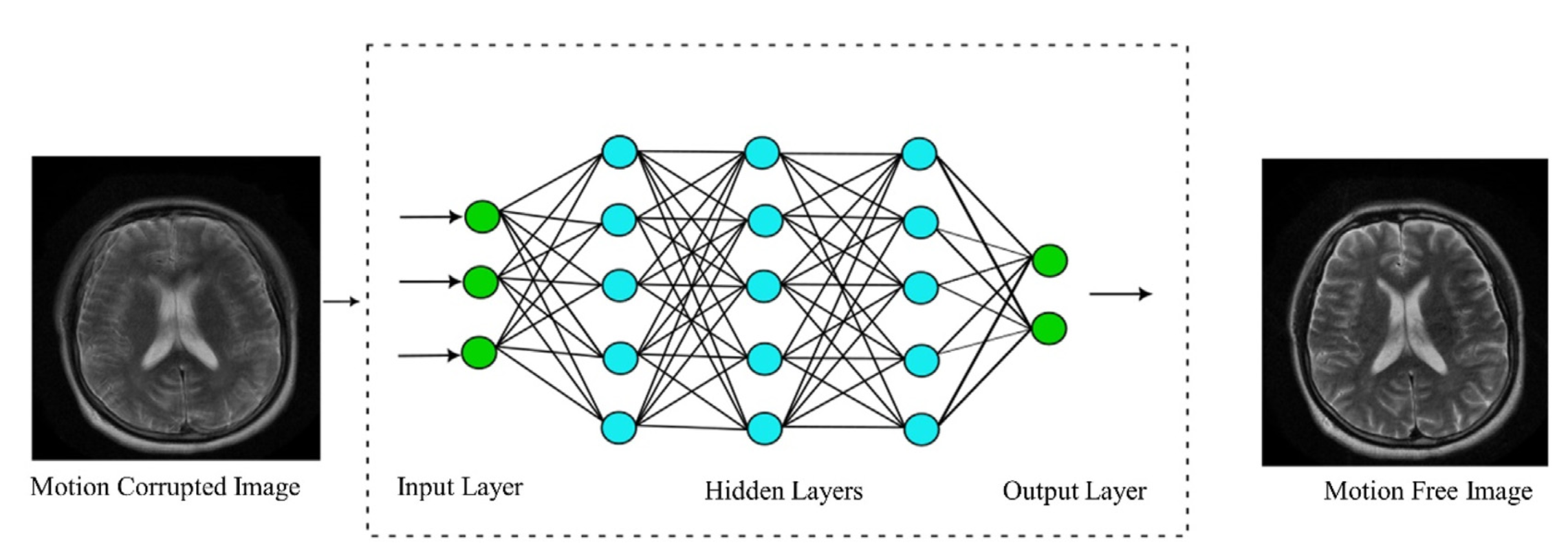}
	\caption{
A flowchart depicting CNN-based motion correction techniques from the work in \textcite{chang2023deep}. Initially, motion-corrupted images are input into a network, which then outputs motion-free images. For training the deep learning model, numerous pairs of motion-corrupted and motion-free images are processed through the network. Subsequently, in the correction phase, a motion-corrected image is produced by inputting a motion-corrupted image. This process ultimately results in the generation of a motion-free image, see the original paper at \href{https://www.sciencedirect.com/science/article/pii/S2950162823000012}{https://www.sciencedirect.com/science/article/pii/S2950162823000012}.}
	\label{fig:DL Motion Correction}
\end{figure*}

Deep Learning-Based Motion Correction methods treat the entire MRI scanning procedure as an input-output system, as depicted in Figure \ref{fig:DL Motion Correction}. Typically, these methods operate within the image domain, where the training process involves using a motion-corrupted image as input and a motion-free image as the target label. Crucially, since the characteristics of motion artifacts vary depending on the motion type, most studies narrow their focus to specific types of motion \autocite{lee2020deep}. During the training phase, these techniques are designed to recognize patterns between images affected by motion and those free from motion artifacts. When encountering a new, motion-distorted image, the system aims to predict and produce a corrected version \autocite{chang2023deep}. Such methods usually concentrate on particular motion artifact types, predominantly categorized into rigid and non-rigid motions. 

Non-rigid or elastic motion, often due to physiological factors, can be classified into two main groups: periodic/continuous motions (such as respiration, cerebrospinal fluid flow, and peristalsis) and abrupt involuntary actions (like swallowing) \autocite{lee2020deep}. 

Conversely, rigid motion is caused by involuntary or deliberate movements of the subject, and is frequently observed across various body parts, especially in the brain \autocite{godenschweger2016motion}. It is more commonly found in non-compliant subjects, including children or individuals with degenerative neurological disorders like Parkinson's \autocite{godenschweger2016motion}. Rigid motion is characterized by six parameters, encompassing three translational and three rotational movements. These motion parameters are inferable directly from the raw data by minimizing motion-induced image quality metrics or data consistency errors. However, applying these methods in clinical MRI settings poses challenges due to the non-convex nature of the parameter estimation process and the extensive computation time involved \autocite{pawar2018moconet}.

\subsection{Convolutional Neural Networks}

In recent years, to increase the captured motion range and accelerate fetal MRI volume reconstruction, neural network-based methods like CNN have been proposed to predict the motion of fetal brain MRI 2D slices. \textcite{miao2016cnn} proposed a CNN-based regression technique, termed Pose Estimation via Hierarchical Learning (PEHL). This method is designed to facilitate real-time 2-D/3-D registration, boasting a broad capture range and high accuracy. While primarily focusing on X-ray attenuation as observed in CT imaging, this approach is also adaptable for MRI applications.

\textcite{salehi2018real} expanded upon the 3D pose estimation network, adapting it for use with arbitrarily oriented objects in slice-to-volume and volume-to-volume registration tasks. This adaptation involved the use of CNNs trained to predict the angle-axis representation of 3-D rotations and translations, based on image features, which is particularly relevant for fetal brain MRI. Meanwhile, \textcite{pei2020anatomy} introduced a multi-task learning framework for fetal MRI stacks. This framework combines the positional data and tissue segmentation maps of each 2D slice, thereby enhancing the effectiveness of motion correction. The framework features a multi-output network designed to simultaneously predict transformation parameters and segmentation results, which are then processed through a common encoder structure and a shared representation module to refine both the predictions and segmentation outcomes \autocite{pei2020anatomy}.

\subsection{Long Short-Term Memory Networks}

\textcite{singh2020deep} firstly proposed a deep predictive motion tracking framework utilizing Long Short-Term Memory (LSTM) Networks. This approach marks a departure from the static 3D pose estimation methods commonly found in \autocite{salehi2018real, miao2016cnn}, focusing instead on dynamic, real-time 3D motion tracking within MRI contexts. In contrast to traditional motion tracking methodologies, their framework directly addresses the 3D rigid motion of anatomy in the scanner or world coordinate system, using time-sequentially acquired slice stacks. 

The process begins with the extraction of spatial features from sequences of input images using Convolutional Neural Networks (CNNs). These features are then encoded using LSTM, which subsequently estimates objectives for the given images. This process results in the creation of a context vector, utilized by LSTM decoders to perform regression against angle-axis representation and translation offset, therefore enabling the accurate prediction of 3D rigid body motion. To prevent overfitting to either rotation or translation parameters, the network incorporates multiple representation heads. Additionally, the framework employs a multi-step prediction strategy, wherein the output of a previous decoder is fed as input to the current decoder, combined with the context vector. The networks were trained and tested on sequences containing masked slices, which represent slices lost due to intermittent fast intra-slice motion.

\subsection{Transformers}

The Transformer architecture, renowned for its proficiency in dynamically highlighting relevant features in input sequences through the self-attention mechanism, has demonstrated exceptional ability in modeling long-distance dependencies and capturing global context \autocite{vaswani2017attention}. 

Formally, the transformer uses the concept of self-attention \autocite{vaswani2017attention}, which allows each position in the input sequence to attend to all positions in the previous layer of the sequence. This is mathematically represented as:
\begin{equation}
\label{eq:attention}
\text{Attention}(Q, K, V) = \text{softmax}\left(\frac{QK^T}{\sqrt{d_k}}\right)V   
\end{equation}
In this equation, $Q$, $K$, and $V$ represent the queries, keys, and values, respectively. These are all vectors obtained by transforming the input vectors. $d_k$ is the dimension of the key vectors. The scaling factor $\sqrt{d_k}$ is used to prevent the dot products from growing too large in magnitude, which could lead to instability in the softmax function.

The output of the attention mechanism is then passed through a series of feed-forward neural networks. Each of these networks is applied to each position separately and identically. This part of the transformer is represented as:

\begin{equation}
\label{eq:ffn}
\text{FFN}(x) = \max(0, xW_1 + b_1)W_2 + b_2
\end{equation}

where \( x \) is the input to the feed-forward network, \( W_1 \) and \( W_2 \) are weight matrices, and \( b_1 \) and \( b_2 \) are bias vectors. The ReLU activation function (denoted as \( \max(0, \cdot) \)) is applied element-wise. This two-layer feed-forward network is applied to each position’s output from the attention layer, effectively allowing the model to integrate information from different positions of the input sequence.

This attention capability is particularly pertinent in the context of SVR for fetal MRI, where the input comprises multiple stacks of slices. These stacks can be conceptualized as a sequence of images, allowing for the joint processing of multi-view information from different orientations to facilitate the SVR task. \textcite{xu2022svort} introduced the Slice-to-Volume Registration Transformer (SVoRT), which maps multiple stacks of fetal MR slices into a canonical 3D space. This mapping serves as an initialization step for SVR and 3D reconstruction. Building on this, NeSVoR has further enhanced the volumetric reconstruction process by incorporating implicit neural information \autocite{xu2023nesvor}.

\subsection{Generative Adversarial Networks}

CNNs, known for their robust feature extraction capabilities, have made substantial contributions to reducing motion artifacts in MRI imaging. However, a limitation arises in the form of blurred images, a consequence of their strategy to minimize the Euclidean distance between noisy and ground-truth images \autocite{isola2017image}. This approach, while mathematically sound in pixel space, often falls short in terms of visual perception, leading to less-than-optimal image quality \autocite{lim2023motion}.

To overcome these limitations, Generative Adversarial Networks (GANs) have been identified as a promising technological advancement. A recent contribution to this area is the work of \textcite{lim2023motion}, who introduced a novel GAN architecture designed specifically for correcting motion artifacts in fetal MRI images.

GANs function through a dual-network structure, comprising a Generator and a Discriminator. These networks are trained sequentially and interactively. The Discriminator's primary function is to develop a loss function capable of distinguishing between authentic and artificial images. Meanwhile, the Generator, operating as a generative model, aims to produce artificial images so convincingly real that they deceive the Discriminator.

In the context of \textcite{lim2023motion} work, the Generator is tasked with transforming a motion-corrupted image into a version free of motion artifacts. This is achieved by the Generator producing a corrected image, which is then combined with the original motion-corrupted image to create a "fake" image pair. Conversely, a "real" image pair is formed by concatenating a ground truth (motion-free) image with the same original motion-corrupted image. This process of pairing the images, particularly the integration of the motion-corrupted image, serves a crucial role. It provides auxiliary information to the Discriminator, enabling it to more effectively identify and penalize inaccuracies in the Generator's output. 

\subsection{Diffusion Model}
Diffusion models represent a pioneering category in generative modeling, showcasing exceptional efficacy in capturing complex data distributions. Although a recent development in generative learning, these models have proven beneficial across a variety of applications. Based on differences in approaches, they can be broadly categorized into two types: the Variational Perspective \autocite{ho2020denoising, xie2022measurement} and the Score Perspective \autocite{kazerouni2022diffusion, song2019generative, song2020score, levac2023accelerated}. We will examine the models within each category, including DDPMs under the Variational Perspective and NCSNs and SDEs under the Score Perspective.

\subsubsection{Variational Perspective}
The Denoising Diffusion Probabilistic Model (DDPM) \autocite{ho2020denoising} is an innovative approach in the realm of unconditional generative models. It employs two distinct Markov chains: a forward chain that progressively distorts data into noise, and a reverse chain that transforms noise back into data. The forward chain is typically designed to convert any data distribution into a simple, predefined prior distribution (such as a standard Gaussian). Conversely, the reverse chain, powered by deep neural networks, learns to invert the forward chain's transformations. Generation of new data points begins by sampling a random vector from the prior distribution, followed by ancestral sampling through the reverse Markov chain \autocite{koller2009probabilistic}.

Consider a data sample $x_0 \sim q(x_0)$. A forward noising process $p$ generates latent variables $x_1$ through $x_T$ by incrementally adding Gaussian noise at each time step $t$. This process is mathematically defined as:
\begin{equation}
\label{eq:forward}
q(x_t \vert x_{t-1}) = \mathcal{N} \left( x_t; \sqrt{1 - \beta_t} \cdot x_{t-1}, \beta_t \cdot \mathbf{I}\right)
\end{equation}
Here, $T$ denotes the total number of diffusion steps, and $\beta_1, \ldots, \beta_T \in [0, 1)$ represents the variance schedule across these steps. $\mathbf{I}$ is the identity matrix, and $\mathcal{N} (x; \mu, \sigma)$ signifies the normal distribution with mean $\mu$ and covariance $\sigma$.

Building on the aforementioned concepts, the reverse process can be formulated to approximate a sample from $q(x_0)$. Starting with $p(x_T) = \mathcal{N}(x_T; 0, \mathbf{I})$, the reverse process can be parameterized as follows:
\begin{equation}
\label{eq:reverse}
p_{\theta}(x_{0:T}) = p(x_T) \prod_{t=1}^{T} p_{\theta}(x_{t-1} | x_t)
\end{equation}

\subsubsection{Score Perspective}
Models employing the Score Perspective utilize a maximum likelihood-based estimation approach, which involves using the score function of the data's log-likelihood to estimate parameters in diffusion processes. Two notable subcategories within this domain are Noise-conditioned Score Networks (NCSNs) \autocite{song2019generative} and Stochastic Differential Equations (SDEs) \autocite{song2020score}. NCSNs specifically focus on estimating the derivative of the log density function for perturbed data distributions at varying noise levels. Conversely, SDEs represent a broader generalization of these methodologies, encompassing characteristics of both DDPMs and NCSNs.

\subsubsection{Application in Motion Correction}
Although diffusion models have not yet been directly implemented in fetal MRI motion correction, several related studies hold potential for such applications. \textcite{xie2022measurement} developed a measurement-conditioned denoising probabilistic model (MC-DDPM), an extension of DDPM, which demonstrated impressive capabilities in reconstructing under-sampled MR images. Such work could be potentially adapted for fetal brain MRI volumetric reconstruction. \textcite{levac2023accelerated} employed a score-based generative model that calculates the log probability of reconstruction for the motion-corrupted MR images. Both motion parameters and the underlying image are optimized to identify a solution that not only aligns with the data but also possesses a high prior probability as per the generative model's framework.

\section{Conclusions}
The analysis of various methodologies reveals that the issue of motion correction in MRI is inherently complex and can be approached from multiple perspectives. While traditional methods such as SVR have proven effective, deep learning-based approaches have demonstrated superior precision and robustness in handling various types of motion. They excel at modeling the intricate non-linear relationships between motion-corrupted and motion-free images. Notably, generative models like GANs and Diffusion Models are advancing toward more accurate reconstruction of corrupted images. However, they often operate in a 'black-box' manner, raising concerns about their lack of transparency, an issue that warrants further exploration and resolution.

Emerging deep learning models, like Transformers, Generative Adversarial Networks, and Diffusion Models, have shown promise in the field of medical imaging. Despite their initial development for MRI reconstruction enhancement, there is potential for their application in motion correction. Especially for diffusion models, which generate samples through a lengthy Markov chain of diffusion steps, are particularly noteworthy. These innovative deep-learning models are likely to significantly enhance motion correction techniques in future research endeavors.

\printbibliography 


\end{document}